\documentclass[11pt]{article}
\usepackage{latexsym}
\usepackage{amsmath,amsbsy,amssymb}
\usepackage{verbatim}
\usepackage{bm}
\usepackage{cite}

\usepackage{curves}	
\usepackage{epic}
\usepackage{eepic}
\usepackage{epsfig}

\newcommand{\dd}{\mbox{\rm d}}

\newcommand{\dg}{\dagger}

\newcommand{\p}{\partial}

\newcommand{\be}{\begin{equation}}
\newcommand{\bear}{\begin{eqnarray}}
\newcommand{\ear}{\end{eqnarray}}
\newcommand{\ee}{\end{equation}}

\newcommand{\lbl}{\label}
\newcommand{\bi}{\bibitem}
\newcommand{\ci}{\cite}

\newcommand{\vs}{\vspace}
\newcommand{\hs}{\hspace}

\newcommand{\vphi}{\varphi}

\newcommand{\vac}{|0 \rangle}

\newcommand{\om}{\omega}
\newcommand{\Om}{\Omega}

\newcommand{\bp}{{\bm p}}

\newcommand{\bx}{{\bm x}}
\newcommand{\bX}{{\bm X}}
\newcommand{\hbp}{{\hat{\bm p}}}
\newcommand{\hbx}{{\hat{\bm x}}}

\begin{document}

\begin{center}

\

\vs{.6cm}

\baselineskip .7cm

{\bf  \Large A Novel View on Successive Quantizations, Leading to Increasingly More  ``Miraculous'' States} 

\vs{4mm}

\baselineskip .5cm
Matej Pav\v si\v c

Jo\v zef Stefan Institute, Jamova 39,
1000 Ljubljana, Slovenia

e-mail: matej.pavsic@ijs.si

\vs{3mm}
{\bf Abstract}
\end{center}

\baselineskip .43cm
{\footnotesize

A series of successive quantizations is considered, starting with the quantization of a non relativistic or relativistic
point particle: 1) quantization of a particle's position, 2) quantization of wave function,
3) quantization of wave functional. The latter step implies that the wave packet profiles forming the
states of quantum field theory are themselves quantized, which gives new physical states that are configurations of
configurations. In the procedure of quantization, instead of the Schr\"odinger first order equation in time
derivative for complex wave function (or functional), the equivalent second order equation for its real
part was used. In such a way, at each level of quantization, the equation a quantum state satisfies is
just like that of a harmonic oscillator, and wave function(al) is composed in terms of the pair of its canonically
conjugated variables.

\vs{2mm}

}

\baselineskip .50cm

\section{Introduction}

In the passage from the classical to the quantum description of a physical system miraculous things happen.
For instance, position of a particle is no longer fixed by the Newton's law of motion or its relativistic
generalization. Instead of position the dynamics is given in terms of a position dependent wave function
$\psi(t,\bx)$ evolving in time. The observed position of a particle at any time $t$ can be anywhere, the probability of
its occurrence being given by $|\psi|^2$. A question now arises as to whether it makes sense to perform
a further step in the quantization procedure and consider $\psi (t,\bx)$ as a quantity to be quantized.
Indeed, such a `` second quantization''\,\ci{Dirac,Jordan,Fock}, where a state can be represented as a functional $\Phi[\psi]$
(see e.g.,\,\ci{Hatfield}) makes much sense and is one of  the cornerstones of the contemporary physics,
although nowadays not interpreted
in the sense of a quantization of a wave function, but just a classical field. We will present here a fresh and
novel view on second quantization, stressing the fact that the state $\Phi [\psi]$ can be expanded in terms
of the Fock space basis states, the superposition coefficients being multiparticle wave packet
profiles $f(t,\bx_1,\bx_2,...,\bx_r)$.  Each of them satisfies the corresponding Schr\"odinger equation,
which for a given initial data gives a definite complex valued function, whose absolute square gives the
probability density of observing at time $t$ a multiparticle
configuration $(\bx_1,...,\bx_r)$\,\ci{PavsicManifestCovariance,PavsicLocalized}. At this point
we can envisage that the functions $f(t,\bx_1,\bx_2,...,\bx_r)$ are in fact the expectation values of some
quantities which are operators of a more basic quantum theory. So we arrive at the quantization III\footnote{
We do not use the name ``third quantization'', which has a different meaning in the literature. For the same reason we will also
not use the name ``second quantization'', but Quantization II.}, namely
the quantization of the wave packet profiles $f(t,\bx_1,\bx_2,...,\bx_r)$, that is, the quantization of the
wave functional $\Phi[t,\psi(\bx)]$. If quantization III is valid, then the probabilities of observing multiparticle
configurations would be in general different from those calculated within the conventional quantum field theory.
So we arrive at the predictions of novel states that we will call ``miracle states''.

\section{The hierarchy of quantizations}
\subsection{Classical system}

Let us consider a classical system, for instance a nonrelativistic particle, described by the action
\be
  I = \int \dd t \, \left ( \frac{m {\dot x^2}}{2}- V(x)   \right ) .
\lbl{2.1}
\ee
The canonicalley conjugated variables are $(x,p)=(x,m {\dot x})$ and the Hamiltonian is
\be
  h = \frac{p^2}{2m}+ V(x)
\lbl{2.2} 
\ee
If for the potential we take
\be
  V(x)=\frac{k x^2}{2} + V_1 (x),
\lbl{2.3}
\ee
then our system is a perturbuded harmonic oscillator.

In the absence of a perturbation, i.e.,, if $V_1 (x)=0$, the system satisfies the equation of motion
\be
   {\ddot x}+ \om^2 x = 0.
\lbl{2.4}
\ee
By introducing the new variables
\be
  \xi = x~,~~~~~~\eta = \frac{{\dot x}} {\om},
\lbl{2.4a}
\ee
the second order differential equations (\ref{2.4}) can be cast into the following system of
two first order differential equations:
\bear
  && {\dot \xi} = \om \eta \lbl{2.5}\\
  && {\dot \eta} = - \om \xi .
\lbl{2.6}
\ear
The latter system can be derived from the action
\be
  I = \int \dd t \left [ \frac{1}{2} (\eta {\dot \xi}- \xi {\dot \eta}) - \om \xi^2 - \om \eta^2 \right ] .
\lbl{2.7}
\ee
Introducing the complex variable $\chi = \xi + i \eta$, the action (\ref{2.7}) becomes
\be
  I = \int \dd t \left [ \frac{i}{2} (\chi^* {\dot \chi} - \chi {\dot \chi}^*)   - \chi^* \om \chi\right ] .
\lbl{2.8}
\ee
The corresponding equations of motion
\bear
   && ~~~i {\dot \chi}= \om \chi , \lbl{2.9}\\
   && - i {\dot \chi}^*= \om \chi^* ,
\lbl{2.10}
\ear
are just like the Schr\"odinger equation for the complex function $\chi (t)$. Its absolute square is
\be
  \chi^* \chi = \xi^2 + \eta^2 = x^2 + \frac{{\dot x}^2}{\om^2} = x^2 + \frac{p^2}{m \om^2}= A^2 ,
\lbl{2.11}
\ee
where $A$ is the oscillation amplitudse. This expression is proportional to the
Hamiltonian $h = \frac{p^2}{2 m}+ \frac{k x^2}{2}$.

In the case of more than one oscillators, a variable bear an index, say $k=1,2,...,n$, and instead of
(\ref{2.11}) we have
\be
  \chi_k^* \chi_k = A_k^2 .
\lbl{2.12}
\ee
We see that $\chi_k$, satisfying the action principle (\ref{2.8}) is just like a wave function. With the
normalization $\sum_{k=1}^n \chi_k^* \chi_k= 1$, the quantity $\chi_k^2 \chi_k$ is like the
probability of observing the $k$-th oscillator excited. Thus, if we perform a measurement
related to the question ``which oscillator'', the outcome of such a measurement is the occurrence
of one of those $n$ oscillators being excited.

The above reasoning would hold if the $\chi_k$ were indeed a wave function. However, in our case of
the classical harmonic oscillators, $\chi_k$ is not wave function. It merely encodes position $x_k$ and
momentum $p_k = m {\dot x}_k$ of the $k$-th oscillator. But anyway, the similarity of the formalism 
expressed in Eqs.\,(\ref{2.8})--(\ref{2.12}) to the formalism of quantum mechanics is striking. Could it be
that $\chi_k^* \chi_k$ is nevertheless associated with the probability of observing excitation of the
$k$-th oscillator? Yes, if we take into account that our classical description is in fact an approximation
to the quantum theory in which $\chi_k$ (and thus $x_k$ and $p_k$) are expectation values of the
corresponding operators.

\subsection{Quantization I :  Position becomes operator}

 Upon quantization of the classical system (\ref{2.1}), $x$ and $p$ become the operators, satisfying the
 commutation relations\footnote{
 Here the indices $k$, $s$ denote either different spatial Cartesian components and/or different oscillators.}
 \be
   [x^k,p_s]= i {\delta^k}_s~,~~~~[x^k,x^s]=0~,~~~~~[p_k,p_s]=0 .
 \lbl{2.13}
 \ee
 Consequently, the Hamiltonian (\ref{2.2}) also becomes operator. In the $x$-representation in which $x$
 is diagonal and $p_k = - i \p/\p x^k \equiv \p_x$, we have
 \be
   h= - \frac{\p_x^2}{2 m} + V(x).
 \lbl{2.14}
 \ee
 
 Instead of a non relativistic particle (\ref{2.1}) we can as well consider a relativistic particle. Then the Hamilton
 operator is either
 \be
   h = \sqrt{-\p_x^2 + m^2},
 \lbl{2.15}
 \ee
 or
 \be
   h = \alpha^i p_i + \beta m ,
\lbl{2.16}
\ee
where $\alpha^i$,  $i=1,2,3$, are the Dirac matrices, satisfying
$[\alpha_i,\alpha_j]= 2 \delta_{ij}$.

A state can be represented by a wave function $\psi(t,x)$, satisfying the Schr\"odinger equation
\be
  i \frac{\p \psi}{\p t}= h \psi ,
\lbl{2.17}
\ee
which can be derived from the action
\be
  I = \int \dd t \, \dd x \left [ \frac{i}{2} (\psi^* {\dot \psi} - \psi {\dot \psi}^*) - \psi^* h \psi    \right ] .
\lbl{2.17a}
\ee
How a quantum wave function can be directly measured was shown by Lunden et al.\,\ci{Lunden}

In the case of the harmonic oscillator, where $V(x)= \frac{k x^2}{2}$, it is convenient to introduce the creation and
annihilation operators
\be
  a_k = \sqrt{\frac{\om m_k}{2}} \left ( x_k + \frac{i p_k}{m \om_k} \right )~,~~~~~~
  a_k^\dg = \sqrt{\frac{\om m_k}{2}} \left ( x_k - \frac{i p_k}{m \om_k} \right ) ,
\lbl{2.17b}
\ee
which apart from the constant factor $\sqrt{\frac{\om m_k}{2}}$ are the operators corresponding to the classical
variables $\chi_k$, $\chi_k^*$, used in Eqs.\,(\ref{2.8})--(\ref{2.12}). A generic state of  a harmonic
oscillator configuration can then be represented in terms of the creation operators acting on the
vacuum wave function $\psi_0$:
\be
  \psi (x) = \sum_{k,n}^\infty c_{kn} ({a_k^{\dg}})^n \psi_0 (x)
  = \sum_{k_1,k_2,...}\sum_{n=0}^\infty c_{k_1 k_2 ...k_n} a_{k_1}^\dg ...a_{k_n}^\dg \psi_0 (x),
\lbl{2.17c}
\ee
where $n$ determines the $n$-th energy eigenstate and $k$ denotes the oscillator degrees of freedom.

The states of particular interest are coherent states $\psi_\alpha$, which are eigenstates of the annihilation operator,
\be
  a_k \psi_{\alpha_k}= \alpha_{k}\psi_{\alpha_k}.
\lbl{2.17d}
\ee
The expectation value is thus
\be
  \langle a_k \rangle = \alpha_k = \sqrt{\frac{2}{\om_k m_k}} \chi_k ~,~~~~~~\chi_k = \langle x_k \rangle 
   + i \frac{\langle p_k\rangle}{m \om_k }
\lbl{2.17e}
\ee
The classical variables $\chi_k$, $\chi_k^*$, used in the previous subsection are thus proportional to
the expectation values of the annihilation operators $a_k$.

Another states of our interest here are superposition of 1st excited states
\be
    \psi = \sum_k c_k a_k^\dg \psi_0 .
\lbl{2.17f}
\ee
Using the expression
\be
  h=  \sum_k \frac{1}{2 m_k}  \left(    - \p_{x_k}^2  + \om_k^2 x_k^2 \right) 
  = \sum_k \om_k \left( a_k^\dg a_k  + \frac{1}{2} \right),
\lbl{2.17g}
\ee
we find from the Schr\"odinger equation (\ref{2.17}) that the superposition coefficients $c_k$
satisfy
\be
  i \frac{\p c_k}{\p t} = \left( \om_k + \frac{1}{2} \right) c_k ,
\lbl{2.17h}
\ee
which, apart from the zero point frequency, is the same as the equation (\ref{2.9}) for $\chi_k$.
The classical variables $\chi_k$ satisfying Eqs.\,(\ref{2.8})--(\ref{2.10}) have indeed their
quantum analogs in $c_k$ whose absolute square $|c_k|^2 = c_k^* c_k$ gives the probability
of observing the excitation of the $k$-th degree of freedom of the system of harmonic
oscillators.

Let us now return to the Schr\"odinger equation (\ref{2.17}), where $h$ is given by (\ref{2.15}),
or its non relativistic approximation (\ref{2.14}) in which we now again take an arbitrary
potential $V(x)$, not necessarily that of a harmonic oscillator. Inserting $\psi = \psi_{R}+ i \psi_{I}$,
Eq.\,(\ref{2.17}) becomes a system of two equations for real functions $\psi_R$ and $\psi_I$:
\be
 {\dot  \psi}_R = h \psi_I ,
\lbl{2.18}
\ee
\be
 {\dot \psi}_I = - h \psi_R ,
\lbl{2.19}
\ee
the corresponding action being 
\be
  I = \int \dd t \, \dd x \left[ \frac{1}{2}(\psi_I {\dot \psi}_I - \psi_R {\dot \psi}_I)  - \psi_R h \psi_R - \psi_I h \psi_I  \right] .
\lbl{2.20}
\ee

The system of two first order equation (\ref{2.18}),(\ref{2.19}) can be rewritten as one second
order equation for the single real function $\psi_R$ which we now rename into $\vphi$:
\be
  {\ddot \vphi}+ h^2 \vphi = 0 ~,~~~~~~~ \vphi \equiv \psi_R .
\lbl{2.21}
\ee
If $h$ is the Hamilton operator (\ref{2.15}) for a relativistic particle, then Eq.\,(\ref{2.21})
is the Klein-Gordon equation for a {\it real} field $\vphi \equiv \psi_R$, derivable from the
action
\be
  I = \frac{1}{2} \int \dd t\, \dd x \left( {\dot \vphi}^2 - \vphi h^2 \vphi  \right) .
\lbl{2.22}
\ee
The second order equation (\ref{2.21}) for a real function $\vphi$ is equivalent to the
Schr\"odinger equation (\ref{2.17}). The real and imaginary part of the wave function
are given by
\be
  \psi_R = \vphi~,~~~~~~~~ \psi_I = h^{-1} {\dot \vphi}.
\lbl{2.23}
\ee
A similar procedure for the case of the non relativistic Hamiltonian (\ref{2.15}) has been considered 
by Deriglazov\,\ci{Deriglazov}.

We see that the relation between the real wave function $\vphi (t,\bx)$ and the
complex-valued wave function $\psi (t,\bx)$ is analogous to the relation between the real variables
$x_k (t)$ and the complex variables $\chi_k (t)$ of the classical harmonic oscillator.
As a classical harmonic oscillator can be formulated either in terms of real $x_k$ or
complex $\chi_k$, satisfying the second order equation (\ref{2.4}) or the first order equation
(\ref{2.9}), respectively, so a quantum system can be formulated either in terms of a
complex wave function $\psi$, or real wave function $\vphi \equiv \psi_R$, satisfying
equations (\ref{2.17}) or (\ref{2.21}),  respectively.

In the literature it is often considered as enigmatic why quantum theory needs complex
quantities. We have shown that quantum mechanics can work with real quantities as
well. Instead of the first order Schr\"odinger equation for complex wave function
we can use the second order equation for a real function $\vphi$, satisfying the
action principle (\ref{2.22}).

Moreover, if $h^2 = m^2 - \p_\bx^2$, then (\ref{2.22}) gives the Klein-Gordon equation.
Interpreting the Klein-Gordon equation as a relativistic equation for a wave function
has caused much problems in the literature. But our analysis here reveals that a real
$\vphi$, satisfying the Klein-Gordon equation is related to a complex wave function
$\psi=\psi_R + i \psi_I$ according to the prescription (\ref{2.23}). The probability
density of observing the particle at position $\bx$ in a given reference frame is
determined by $|\psi (t,\bx)|^2$. The seeming intricacies concerning Lorentz covariance
and other related concerns were clarified in \ci{PavsicManifestCovariance,PavsicLocalized},
and references therein.

Using (\ref{2.23}), we obtain for the probability density the expression
\be
  \psi^* \psi = \vphi^2 + h^{-1} {\dot \vphi}h^{-1} {\dot \vphi} ,
\lbl{2.24}
\ee
which is analogous to Eq.\,(\ref{2.11}) of the classical harmonic oscillator.
Here $h$ is the non relativistic (\ref{2.14}) or the relativistc (\ref{2.15}) Hamilton
operator. It can also be the Dirac Hamilton operator (\ref{2.16}). Namely, in the
presence of a complex-valued Hamiltonian (\ref{2.16}), $h=h_r + i h_I$,
we have
\bear
  &&{\dot \psi}_R = h_R \psi_I + h_I \psi_R ,  \lbl{2.25}\\
  &&{\dot \psi}_I = - h_R \psi_R + h_I \psi_I ,  \lbl{2.26}
\ear
which gives
\be
  {\ddot \psi}_R + \left(  h_R^2 - h_I^2 \right) \psi_R = 0.
\lbl{2.27}
\ee
Using the anticommutativity of the matrices $\alpha^i = \alpha_R^i + i \alpha_I^i$,
we find $h_R h_I + h_I h_R =0$ and therefore $h^2 = (h_R + i h_I)^2 = h_R^2 - h_I^2$.
Thus the equation (\ref{2.27}) has the same form as (\ref{2.21}). It is a second order
equation for a real 4-component object $\psi_R$.

A real 1-component field $\vphi$, satisfying the Klein-Gordon equation, derived from the
action (\ref{2.21}), cannot describe antiparticles, neither charged nor neutral. For
description of charged particles one needs a 2-component field. For description of
charged spin 1/2  particles one needs a real 4-component field $\vphi \equiv \vphi_s$,
$s=1,2,3,4$, satisfying the Klein-Gordon equation (\ref{2.27}) for each component
$\vphi \equiv \psi_R$. As we have shown above, the latter equation can be written
in the form of the first order equation (\ref{2.17}) for a complex wave function, in
which $h$ is now the Dirac Hamiltonian (\ref{2.16}). In other words, the 2nd order
Klein-Gordon equation for a 4-component real scalar field can be written in the
form of the Dirac equation for a complex 4-component field $\psi$.

The passage from the Klein-Gordon equation to the system of 1st order equations can
be done as well by introducing the new variables
\bear
   &&\chi_1 = \left( i \frac{\p}{\p t} + h  \right) \vphi  ,  \hs{2cm}\lbl{2.28}\\
   &&\chi_2 = \left( -i \frac{\p}{\p t} + h  \right) \vphi,  \hs{2cm}\lbl{2.29}
\ear
which satisfy
\bear
   &&\left( -i \frac{\p}{\p t} + h  \right) \chi_1 = 0, \hs{2cm}  \lbl{2.30}\\
   &&\left( i \frac{\p}{\p t} + h  \right) \chi_2 = 0,  \hs{2cm}  \lbl{2.31}
\ear
Such a procedure is usually considered in the literature, following Foldy\,\ci{Foldy}.
We wish to point out here that for a real scalar field $\chi_2 = \chi_1^*$, and therefore
equations (\ref{2.30}),(\ref{2.31}) are not independent. Instead of the Klein-Gordon
equation we have the Schr\"odinger equation for a complex wave function $\chi \equiv \chi_1$.
From Eqs.\,(\ref{2.28}) and (\ref{2.23}) we find the relation
\be
  \psi = \psi_R + i \psi_I = \left(  \vphi + i h^{-1} {\dot \vphi} \right) = h^{-1}\chi .
\lbl{2.32}
\ee
If $\vphi$ is a 4-component real field, then $\chi_1 \equiv \chi$ is a 4-component complex wave
function, and if the Hamiltonian $h$ is given Eq.\,(\ref{2.16}), then (\ref{2.30}) is the
Dirac equation, while (\ref{2.31}) is its cpmplex conjugate.

Important lesson that we learned here is that although a real or complex function $\vphi (t,\bx)$,
satisfying the Klein-Gordon equation is not a wave function that gives the probability
density of finding a particle at position $\bx$, one obtains such a wave function as
a superposition of $\vphi (t,\bx)$ and its  conjugated momentum
$p_\vphi = {\dot \vphi}$. The corresponding conserved probability current density is obtained
upon differentiation of $\psi^* \psi$ with respect to time and taking into account the Schr\"odinger
equation with a particular Hamilton operator, such as (\ref{2.14}), (\ref{2.15}) or (\ref{2.16}).
How the relativistic probability density and the probability current can be treated in a Lorentz covariant
way is shown in Ref.\,\ci{PavsicManifestCovariance}.

\subsection{Quantization II : Wave function becomes operator}

A quantum state is usually represented by a complex wave function
$\psi (t,\bx) = \psi_R +i \psi_I$ satisfying the action principle (\ref{2.17a}) leading to the
first order Schr\"odinger equation (\ref{2.17}). We have seen that a quantum state can
as well be represented by a real function $\vphi= \psi_R$, satisfying the action principle
(\ref{2.22}) which gives the second order equation (\ref{2.21}), known as the Klein-Gordon
equation. The Hamiltonian $h$ can be given by Eq.\,(\ref{2.14}), (\ref{2.15}) or (\ref{2.16}).

At this point we can envisage that $\vphi (t,\bx)$, satisfying the Klein-Gordon equation,
is a field that corresponds to the expectation
value of the corresponding operator. In other words, we are now quantizing the field
$\vphi(t,\bx)$. Instead of the harmonic oscillator like action (\ref{2.22}), let us now take a
more general action
\be
I = \frac{1}{2} \int \dd t \,  \dd^3 \bx \left( {\dot \vphi}^2 - \vphi h^2 \vphi + V_1 (\vphi) \right) 
\lbl{2.33}
\ee
that contains an interaction term $V_1 (\vphi)$. The canonically conjugated variables
are $\vphi$ and $p_\vphi = {\dot \vphi}$ which upon quantization become operators, $\hat \vphi$,
${\hat p}_\vphi$, that in the Schr\"odinger picture can be represented as
\be
     {\hat \vphi} \longrightarrow \vphi(\bx)~,~~~~~~~~{
     \hat p}_\vphi \longrightarrow - i \frac{\delta}{\delta \vphi(\bx)}.
\lbl{2.33a}
\ee

The Hamilton operator is
\be
  H = \frac{1}{2} \int \dd^3  \bx \left( \frac{\delta^2}{\delta \vphi^2}+ \vphi h^2 \vphi + V_1 (\vphi)    \right) .
\lbl{2.33b}
\ee
Introducing the operators\footnote{Notice that $a(\bx)= \frac{\sqrt{h}}{\sqrt{2}}
    	\left(  \vphi + i h^{-1} p_\vphi \right)$ corresponds to the wave function 
$\psi =  \vphi + i h^{-1} p_\vphi$ of the unquantized theory.}
\bear
  &&a(\bx)=\frac{1}{\sqrt{2}} \left(  \sqrt{h}\vphi + i h^{-1/2} p_\vphi \right), \lbl{2.34}\\
  &&a^\dg(\bx)=\frac{1}{\sqrt{2}} \left(  \sqrt{h}\vphi - i h^{-1/2} p_\vphi \right), \lbl{2.35}
\ear
which satisfy
\be
  [a(\bx),a^\dg (\bx')] = \delta^3 (\bx-\bx')~,~~~[a(\bx),a(\bx')] = 0~,~~~[a^\dg(\bx),a^\dg(\bx')] = 0,
\lbl{2.35a}
\ee
we have
\be
  H = \frac{1}{2} \int \dd^3 \bx \left( a^\dg (\bx)\,h a(\bx) + a(\bx)h a^\dg (\bx) + V_1 \left( \frac{1}{\sqrt{2 h}}
  (a(\bx) +a^\dg(\bx))   \right)      \right) .
\lbl{2.35b}
\ee

A generic state can be represented as a functional
\be
  \Phi[\vphi]= \sum_{r=1}^\infty \dd^3 \bx_1\dd^3 \bx_2 ... \dd^3 \bx_r f(t,\bx_1,\bx_2,...,\bx_r) 
  a^\dg (\bx_1)...a^\dg (\bx_r) \Phi_0[\vphi],
\lbl{2.36}
\ee
satisfying the Schr\"odinger equation
\be
   i \frac{\p \Phi}{\p t} = H \Phi .
\lbl{2.37}
\ee

The functional $\Phi[\vphi]$ is a superposition of the multiparticle basis states generated by the
action of the creation operators (\ref{2.35}) on the vacuum functional that satisfies
$a(\bx)\Phi_0 [\vphi]=0$. The superposition coefficients are multiparticle wave functions
(wave packet profiles) $f(t,\bx_1,\bx_2,...,\bx_r)$. Inserting the expression (\ref{2.36}) into the
Schr\"odinger equation (\ref{2.37}), we find, in the case when $V_1[\vphi]=0$, that
\be
  i \frac{\p f(t,\bx_1,\bx_2,...,\bx_r)}{\p t} = \sum_{k=1}^r \om_{\bx_k}f(t,\bx_1,\bx_2,...,\bx_k),
\lbl{2.38}
\ee
where, depending on the system under consideration\footnote{
In the case of the Dirac Hamiltonian, the commutators in Eq.\,(\ref{2.35a}) must be replaced by anticommutators.},
 $\om_{\bx_k}= - \frac{\p_{\bx_k}^2}{2 m} + V(\bx_k)$,
or $\om_{\bx_k}= \sqrt{m^2 - \p_{\bx_k}^2}$, or $\om_{\bx_k} = i \alpha^i \p_{\bx_k} + \beta m$.
Here $\p_{\bx_k} \equiv \frac{\p}{\p \bx_k}\equiv \frac{\p}{\p \bx_k} \equiv \nabla$.
If $V_1 [\vphi]\neq 0$, then (Eq.\,\ref{2.38}) contains additional terms that mix multiparticle configurations.
Then
\be
i \frac{\p f^{(\bx_1,\bx_2,...,\bx_r)}}{\p t}
= \sum_{s=1}^\infty C_{~~~~(\bx'_1,\bx'_2,...,\bx'_s)}^{(\bx_1,\bx_2,...,\bx_r)} f^{(\bx'_1,\bx'_2,...,\bx'_s)},
\lbl{2.39}
\ee
where the integration is assumed over dummy continuous indices $(\bx'_1,\bx'_2,...,\bx'_s)$.
For our purpose here it is not necessary that we present explicit expressions for the coefficients
$C_{~~~~(\bx'_1,\bx'_2,...,\bx'_s)}^{(\bx_1,\bx_2,...,\bx_r)}$. In the case of single particle wave function,
if $V_1 {\vphi}=0$, it is
\be
i \frac{\p f^{(\bx)}}{\p t}
= C_{~(\bx')}^{(\bx)} f^{(\bx')} ~,~~~~~~~~~C_{~(\bx')}^{(\bx)} = h \delta_{~~{(\bx')}}^{(\bx)} ,
\lbl{2.40}
\ee
where $\delta_{~~{(\bx')}}^{(\bx)} \equiv \delta^3 (\bx - \bx')$, and $h$ is one of the 1st quantized Hamiltonins
given in Eqs.\, (\ref{2.14})--(\ref{2.16}).

A task of quantum field theoretic calculations is to find explicit expressions for the wave packet profiles
evolving in time. Their evolution is fixed if initial values are known. Once $f(t=0,\bx_1,...,\bx_r)$ are known,
also the wave functional is fixed; it is fixed by the Schr\"odinger equation (\ref{2.37}).
In such theory the wave functional $\Phi[\vphi]$ obeys a deterministic law, namely (\ref{2.37}).
The absolute square $|\Phi[\vphi]|^2$ gives the probability density of observing a field\footnote{
For simplicity we use the same symbol for a field operator ${\hat \vphi}(\bx)$ and for its eigenvalues
$\vphi(\bx)$, determined according to ${\hat \vphi}(\bx)|\vphi(\bx) \rangle = \vphi(\bx) |\vphi(\bx) \rangle$.
Recall that we use the representation (\ref{2.33a}).  } $\vphi(\bx)$.

The expectation values are:
\be
  \langle {\hat \vphi} \rangle = \int {\cal D}\vphi\, \Phi^* \vphi \Phi~,~~~~
  \langle {\hat p}_\vphi \rangle = \int {\cal D} \vphi\, \Phi^* \left( - i \frac{\p}{\p \vphi}  \right) \Phi,
\lbl{2.41}
\ee
\be
  \frac{\dd}{\dd t} \langle {\hat \vphi} \rangle = \int {\cal D} \vphi \left( \frac{\dd \Phi^*}{\dd t} \vphi \Phi
  + \Phi^* \vphi  \frac{\dd \Phi}{\dd t}  \right)
  = \int {\cal D} \vphi \, \Phi^* (-i)[\vphi,H] \Phi = \int {\cal D}\, \Phi^* p_\vphi \Phi
\lbl{2.42}
\ee
\be
   \frac{\dd}{\dd t} \langle {\hat p}_\vphi \rangle =  \int {\cal D} \vphi \, \Phi^* (-i)[p_\vphi,H] \Phi
   = \int {\cal D} \vphi \Phi^* (-h^2 \vphi - \frac{\p V_1}{\p \vphi}) \Phi =
    - h^2  \langle \vphi \rangle - \langle \frac{\p V_1}{\p \vphi} \rangle .
\lbl{2.43}
\ee
Because $p_\vphi = {\dot \vphi}$, we have thus confirmed that the expectation values satisfy the
classical field equation, derived from the action (\ref{2.33}).

Let us now consider the operator
\be
  {\hat \bx} = \int \dd^3 \bx\, a^\dg (\bx)\, \bx \, a(\bx) .
\lbl{2.44}
\ee
Acting on a single particle state, the latter operator gives
\be
  \hbx \, a^\dg (\bx) \Phi_0 = \bx \, a^\dg (\bx) \Phi_0 .
\lbl{2.45} 
\ee
In a non relativistic theory, employing the Hamiltonian (\ref{2.14}), the role of the operator
$\hbx$ as position operator is unquestionable. In Refs.\,\ci{PavsicLocalized,PavsicManifestCovariance}, it has
been shown that also in a relativistic theory, based on the Hamiltonian (\ref{2.15}),
the expression (\ref{2.44}) makes sense as the position operator. Anyway, the arguments presented in
this paper do not depend on the usage of the $\bx$-representation and the operator (\ref{2.44}). One could
as well use the momentum representation and the momentum operator $\bp = \int \dd^3 \bp\, a^\dg (\bp)\, \bp \,a (\bp)$
and momentum space wave packet profiles $g(t,\bp_1,\bp_2,...)$.

If we calculate the expectation value in a generic state (\ref{2.36}), we find
  $$\langle \hbx \rangle = \int \dd^3 \bx \, f^* (\bx)\, \bx f (\bx) + \int \dd^3 \bx_1 \,\dd^3 \bx_2 f^*(\bx_1,\bx_2)
   (\bx_1 + \bx_2) f(\bx_1,\bx_2)+ ...$$
\be   
    \hs{1cm}=\sum_{r=1}^\infty \int \dd^3 \bx_1 \,\dd^3 \bx_2...\dd \bx_r  f^*(\bx_1,\bx_2,..., \bx_r)
   \left (\sum_{s=1}^r \bx_s  \right )f(\bx_1,\bx_2,..., \bx_r)
\lbl{2.46}
\ee
We see that in quantization II, in which the wave function is quantized, the expectation value for the
operator $\hbx$ in general deviates from the expectation $\langle \hbx \rangle = \int \dd^3 \bx \, f^* (\bx)\, \bx f (\bx)$,
calculated according to quantization I, where the wave function $\psi(t,\bx)$ corresponds to the
single particle wave packet profile $f(t,\bx)$. Namely, besides single particle wave packet profiles there
are in general multiparticle wave packet profiles, which also contribute to the expectation value
$\langle \hbx \rangle $.

\subsection{Quantization III: Wave functional becomes operator}

A next step is to quantize the wave functional $\Phi[\vphi]$ that satisfies the Schr\"odinger
equation (\ref{2.37}). Folowing the same procedure as in Sec.\,2.2, we now write 
$\Phi [\vphi] = \Phi_R + i \Phi_I$, identify $\Phi_R [\vphi] \equiv \chi [\vphi]$, and arrive
at the equation, equivalent to (\ref{2.37}),
\be
  {\ddot \chi} + H^2 \chi = 0 ~,~~~~~~~~~\chi [\vphi]\equiv \Phi_R[\vphi],
\lbl{2.47}
\ee
where $H$ is given by Eq.\,(\ref{2.33b}).

The corresponding action is
\be
  I = \int \dd t {\cal D} \vphi \, \frac{1}{2}\left( {\dot \chi}^2+ \chi H^2 \chi  \right) ,
\lbl{2.48}
\ee
and the Hamiltonian
\be
  {\cal H}= \frac{1}{2} \int {\cal D} \vphi \left( P_\chi^2 + \chi H^2 \chi   \right) ,
\lbl{2.48a}
\ee
where ${\cal D} \vphi$ stands for the measure of the functional integral in the space of fields $\vphi(\bx)$
This is now the ``classical'' system to be quantized. The canonically conjugated variables
are $\chi$ and $P_\chi = {\dot \chi}$. Again, the action is just like the harmonic oscillator action
for the variable $\chi$.

Into the action (\ref{2.48}) we can as well include an interaction term ${\cal V}_1 [\chi]$, so
that the Hamiltonian becomes
\be
  {\cal H }= \frac{1}{2} \int {\cal D} \vphi \left( P_\chi^2 + \chi H^2 \chi +  {\cal V}_1 [\chi]  \right) ,
\lbl{2.49}
\ee
in which case Eq.\,(\ref{2.47}) is replaced by
\be
 {\ddot \chi} + H^2 \chi + \delta {\cal V}_1/\delta \chi =0.
\lbl{2.47a}
\ee

Upon quantization $\chi$ and $P_\chi$ become operators that can be represented as
\be
  \chi \rightarrow \chi[\vphi] ~,~~~~~~ P_\chi \rightarrow - i \frac{\delta}{\delta \chi [\vphi]} .
\lbl{2.50}
\ee

A state can be represented as a functional $\Psi[\chi[\vphi]]$ of a functional $\chi[\vphi]$.
It satisfies the Schr\"odinger equations
\be
  i \frac{\p \Psi}{\p t} = {\cal H} \Psi,
\lbl{2.51}
\ee
where the Hamilton operator is now (\ref{2.49}).

Introducing the operators
\bear
  &&A[\vphi]= \frac{1}{2} \left( H^{1/2} \chi + i H^{-1/2} P_\chi   \right) , \lbl{2.52} \\
  &&A^\dg\vphi]= \frac{1}{2} \left( H^{1/2}\chi -  i H^{-1/2} P_\chi   \right) , \lbl{2.53},
\ear
satisfying
\be
  [A[\vphi],A^\dg [\vphi']] = \delta (\vphi - \vphi') ~,~~~~ [A[\vphi],A] [\vphi'] = 0 ~,
  ~~~~~ [A^\dg[\vphi],A^\dg [\vphi'] ]= 0,
\lbl{2.53a}
\ee
the Hamiltonian (\ref{2.49}) becomes
\be
    {\cal H }=  \int {\cal D} \vphi \left( \frac{1}{2}\left(   A^\dg[\vphi]H A[\vphi]  +  A[\vphi] H A^\dg[\vphi] \right)  +{\cal V}_1 [A,A^\dg]  \right) ..
\lbl{2.54}
\ee
A generic state, satisfying the Schr\"odinger equation (\ref{2.51}), can be written as
\be
  \Psi[\chi] = \sum_r {\cal D}\vphi_1 ... {\cal \vphi}_r F[\vphi_1,...,\vphi_r] 
  A^\dg [\vphi_1].. A^\dg [\vphi_r] \Psi_0[\chi] , 
\lbl{2.55}
\ee
where $ \Psi_0[\chi]$ is vacuum functional, that is annihilated according to $A[\vphi]  \Psi_0[\chi] =0$,
and $F[\vphi_1,.\vphi_2,..,\vphi_r]$ are complex valued ``wave packet'' profiles. The expression (\ref{2.55}) is
analogous to the expression (\ref{2.36}) which, in turn, is analogous to the expression (\ref{2.17c}) for
a harmonic oscillator.

The operator $A^\dg [\vphi]$ acting on the vacuum creates a field configuration $\vphi(x)$. The wave packet
$F[\vphi]$ gives the probability amplitude of observing a field configuration $\vphi(x)$, the probability
density being $|F[\vphi]|^2$.  Reapeated action of $A$'s gives multi field states
$A^\dg [\vphi_1].. A^\dg [\vphi_r] \Psi_0[\chi]$, where in particular, the fields can be equal\footnote{
   We are now considereing a bosonic field. For a fermionic field the commutator in (\ref{2.53a}) should
   be replaced by anticommutators.},
$\vphi_1 = \vphi_2 = ... \vphi_r$, in which case we have excited states for a single field.

If we insert the expression (\ref{2.55}) into the Schr\"odinger equation (\ref{2.51}), we obtain a series of equations
for $F[\vphi_1,...,\vphi_r]$, $r=1,2,..., \infty$, analogous to (\ref{2.38}) if ${\cal H}_1=0$,
and (\ref{2.39}) if ${\cal H}_1 \neq 0$. In analogy to the single particle
(\ref{2.40}) we have now for ${\cal H}_1=0$ the equation
\be
  i \frac{\p F[\vphi]}{\p t} = H F[\vphi] .
\lbl{2.56}
\ee

The above equation is a thus a consequence of the Schr\"odinger equation (\ref{2.51}) if ${\cal H}_1 =0$.  
It is the same as the Schr\"odinger equation (\ref{2.37}) of Quantization II, and $F[\vphi]$ can be identified
with the wave functional $\Phi[\vphi]$. Quantization II is thus a special case of Quantization III, confined
to the single field case in the expansion (\ref{2.55}) of $\Psi[\chi]$ in term of multi field states.

For the expectation values of the operators ${\hat \chi}$ and ${\hat P}_\chi$ we have
\be
  \langle {\hat \chi} \rangle = \int {\cal D}\chi \,\Psi^* {\hat \chi} \Psi ~,~~~~~
  \langle {\hat P}_\chi \rangle = \int {\cal D} \chi \,\Psi^* \left(   - i \frac{\delta}{\delta \chi}  \right) \Psi ,
\lbl{2.57}
\ee
\be
\frac{\dd \langle {\hat \chi} \rangle}{\dd t}= \int {\cal D}\chi \,\Psi^* (-i) [{\hat \chi}, {\cal H}] \Psi
= \int {\cal D} \chi \,\Psi^* {\hat P}_\chi \Psi  =  \langle {\hat P}_\chi \rangle ,
\lbl{2.58}
\ee
\be
  \frac{\dd \langle {\hat P}_\chi \rangle}{\dd t}= \int {\cal D} \chi \,\Psi^* (-i) [{\hat P}_\chi, {\cal H}] \Psi
  = \int {\cal D} \chi \,\Psi^* \left( - H^2 \chi - \frac{\delta {\cal V}_1}{\delta \chi}  \right)
  = - H^2  \langle {\hat \chi} \rangle  = \langle \frac{\delta {\cal V}_1}{\delta \chi} \rangle .
\lbl{2.59}
\ee
The expectation values thus satisfy the equation (\ref{2.47a}) for the ``classical '' variable
$\chi[\vphi]$, before it is promoted to operator.

Quantizing $\chi[\vphi]$, satisfying the Klein-Gordon like equation (\ref{2.47}), is equivalent to
quantizing $\Phi[\vphi]$, satisfying the Schr\"odinger equation (\ref{2.37}). Because $\Phi[\vphi]$
can be expanded according to (\ref{2.36}), its quantization means that the wave packet profiles
$f(\bx_1,\bx_2,...,\bx_r)$, $r=1,2,...,\infty$, become operators. They satisfy the action principle\footnote{
Instead of the action principle for complex  $f = f_R + i f_I$, we can as well use the action principle
for the real function $f_R$, satisfying ${\ddot f} (\bx) + h^2 f_R (\bx)=0$, and the corresponding multiparticlce
equations for $f_R (\bx_1,\bx_2,...,\bx_r)$.    }
$$
  I =  \int \dd t \,\dd^3 \bx \left( i f^* {\dot f}- f^* \om_\bx f     \right)
  +  \int \dd^3 \bx_1 \dd^3 \bx_2 \left (i f^*(\bx_1,\bx_2) {\dot f}(\bx_1,\bx_2) \right . $$
\be
  \hs{2cm} -  f^* (\bx_1,\bx_2) \om_{\bx_1} f(\bx_1,\bx_2)  -   f^* (\bx_1,\bx_2) \om_{\bx_2} f(\bx_1,\bx_2) \Bigr) + ... ,
\lbl{2.60}
\ee
which gives the series of equations (\ref{2.38}).

Using the compact notation $(\bx_1,\bx_2,...,\bx_r) \equiv {\bm X}_r$, $~\dd \bx_1 \dd \bx_2 ...\dd \bx_r 
\equiv \dd \bX_r$, the action (\ref{2.60}) becomes 
\be
  I = \sum_r  \int \dd t \, \dd \bX_r \left( i f^* (\bX_r) {\dot f} (\bX_r) - f^* (\bX_r) \Omega_{\bX_r} f(\bX_r)     \right),
\lbl{2.61}
\ee
where $\Omega_{\bX_r} = \sum_{k=1}^r \om_{\bx_r}$. Introducing the notation
$\Om_{(\bX_r)(\bX_s)}\equiv   \Omega (\bX_r,\bX_s) = \sum_{k=1}^r \om_{\bx_k} \delta (\bX_r -\delta \bX_s)$, the latter action can be
written as
$$
  I = \int \dd t  \left ( \sum_r \int \dd \bX_r f^*(\bX_r) {\dot f }(\bX_r) - \sum_{r s} \int \dd \bX_r \dd \bX_s
  f^* (\bX_r) \Om(\bX_r,\bX_s) f(\bX_s)  \right )$$
  \be
    =
  \int \dd t \left ( f^{*(\bX_r)} {\dot f}_{(\bX_r)} - f^{*(\bX_r)} \Om_{(\bX_r)(\bX_s)} f^{(\bX_s)} \right ),
\lbl{2.61a}
\ee
where the repeated up and down indices imply the integration over $\dd \bX_r$ and the summation over $r$.
The Hamiltonian is then
\be
   H = f^{*(\bX_r)} \Om_{(\bX_r)(\bX_s)} f^{(\bX_s)}.
\lbl{2.61b}
\ee

Upon quantization, the wave packet profiles are no longer given deterministacally.  Instead there exists a
probability density of observing a definite $f(\bx_1,...,\bx_r)\equiv f(\bX_r)$, given by the absolute square,
$|\theta[f[(X_r)]]|^2$, of the probability amplitude, which is a functional $\Theta[f(X_r)]$.
Wave packets $f(X_r)$ and its conjugate momenta are $f^*(X_r)$ become operators 
\be
  {f}(X_r) \longrightarrow {\hat f}(X_r) \equiv A(\bX_r)~,~~~~~~
  { f}^*(X_r) \longrightarrow {\hat f}^* (\bX_r) \equiv A^\dg (\bX_r) ,
\lbl{2.62}
\ee
satisfying
\be
  [A(X_r),A^\dg (X'_r)] = \delta(X_r - X'_r),~~~[A(X_r),A^ (X'_r)] =0,~~~[A^\dg(X_r),A^\dg (X'_r)] =0.
\lbl{2.65}
\ee
They annihilate and create multiparticle configurations, vacuum being defined according
to $A (\bX_r) \vac =0$. The above commutation relations imply that the eigenvalues of
${\hat f}(\bX_r)$ and ${\hat f}^* (\bX_r)$ cannot be simultaneously determined. 

Defining vacuum state according to $A(\bX_r) \vac = 0$, a generic state can be expressed as
\be
  |\Psi \rangle= \sum_k \dd \bX_{r_1}\dd \bX_{r_2}...\dd \bX_{r_k} \Psi (\bX_{r_1},..., \bX_{r_k})
    A^\dg (\bX_{r_1}) ... A^\dg (\bX_{r_k}) \vac.
\lbl{2.66}
\ee

A particular case of the operators $A[\bX_r]$, $A^\dg[\bX_r]$ for $r=1$ are $a(\bx)$, $a^\dg (\bx)$ that
annihilate and create a single particle, a generic multiparticle state being a superposition (\ref{2.36}) of
the configuration states $|\bx \rangle = a^\dg (\bx) \vac$, $|\bx_1,\bx_2 \rangle = a^\dg (\bx_1) a^\dg (\bx_2)\vac$,
..., etc., or shortly
\be
  |\bX_r \rangle = A^\dg (\bX_r) \vac = a^\dg (\bx_1)... a^\dg (\bx_r )\vac~, ~~~~~r=1,2,.,,,,\infty .
\lbl{2.66a}
\ee
The generic state  (\ref{2.66}) of Quantization III thus contains as as special case the generic state
of Quantization II, represented as the
functional $\Phi[\vphi]$ (Eq.\,(\ref{2.36})). Besides the basis configuration states
$|\bX_r \rangle = |\bx_1,\bx_2,...,\bx_r \rangle$ we now also have multi configuration basis states
\be
  |\bX_{r_1},\bX_{r_2},...\bX_{r_k} \rangle = A^\dg (\bX_{r_1}) A^\dg (\bX_{r_2}) ... A^\dg (\bX_{r_k}) \vac ,
\lbl{2.67}
\ee
that form a basis of a higher order Fock space.

Using now again the notation ${\hat f}(\bX_r)$, ${\hat f}^* (\bX_r)$ for the operators $A(\bX_r)$,
$A^\dg(\bX_r)$, and
representing them according to ${\hat f}(\bX_r) \rightarrow f(\bX_r)$, ${\hat f}^*(\bX_r) \rightarrow 
-i \frac{\delta}{\delta f (\bx_r)}$,
we can calculate their expectation values as
\be
\langle  {\hat f} (\bX_r) \rangle = \int  {\cal D} f \,\Theta [f] \,f(\bX_r) \,\Theta [f] ~,~~~~~
\langle  {\hat f}^* (\bX_r) \rangle = \int  {\cal D} f \, \Theta [f] \left ( -i \frac{\delta}{\delta f}  \right ) \Theta [f] ,
\lbl{2.68}
\ee
where $\Theta [f] = \langle f|\Psi \rangle$.
Analogously as in Eqs.\,(\ref{2.57}--(\ref{2.59}) we can now derive the ``classical'' equation
of motion (\ref{2.38}) as the expectation value equations
\be
  \frac{\dd}{\dd t} \langle {\hat f}^* (\bX_r) \rangle 
  = \int {\cal D} f \,\Theta^* [f] \,(-i)[ f^* (\bx),\Omega_{\bX_r}] \,\Theta [f] .
\lbl{2.69}
\ee

In analogy to the position operator (\ref{2.44}) for a single particle, we can now define
the position operator for a configuration $\bX_r \equiv (\bx_1,\bx_2,...,\bx_r)$:
\be
  {\hat \bX}_r = \int \dd \bX_r \, A^\dg (\bX_r) \bX_r A (\bX_r)~ ,~~~~~r=1,2,3,...,\infty .
\lbl{2.70}
\ee
Explicitly this reads
\bear
   &&\hs{1.7cm} \hbx =\int \dd^3 \bx \, a^\dg (\bx) \bx \, a (\bx), \nonumber \\
   &&\hs{6mm}(\hbx_1,\hbx_2) = \int \dd^3 \bx_1 \dd^3 \bx_2 \, a^\dg (\bx_1) a^\dg (\bx_2 )(\bx_1,\bx_2)
     a (\bx_1) a (\bx_2 ), \nonumber \\
   &&(\hbx_1,\hbx_2,\hbx_3) = \int \dd^3 \bx_1 \dd^3 \bx_2  \dd^3\bx_3\, 
   a^\dg (\bx_1) a^\dg (\bx_2 ) a^\dg (\bx_3) (\bx_1,\bx_2,\bx_3)
     a (\bx_1) a (\bx_2 ) a(\bx_3), \nonumber \\
     &&\hs{3cm}\vdots
\lbl{2.70a}
\ear

The eigenvalues of ${\hat \bX}_r$ in a state with a definite configuration, $|\bX_r \rangle
\equiv |\bx_1,\bx_2,..., \bx_r \rangle \vac $ are
\be
  {\hat \bX}_r |\bX_r \rangle = \bX_r |\bX_r \rangle .
\lbl{2.71}
\ee

The expectation value of a configuration position operator in a generic state (\ref{2.66}) is
\bear
  &&\langle {\hat \bX}_r \rangle = \int {\cal D} f \, \Theta^*[f] {\hat \bX}_r \Theta [f] 
  = \int {\cal D}\, \bX_r \psi^* (\bX_r) \bX_r \psi (\bX_r) \nonumber\\
  &&\hs{2,3cm} + \int {\cal D} \bX_{r_1} {\cal D} \bX_{r_2}\,
  \psi^* (\bX_{r_1},\bX_{r_2}) (\bX_{r_1} + \bX_{r_2}) \psi(\bx_{r_1},\bX_{r_2}) + ... \nonumber \\
  &&\hs{1cm}=\sum_{k=1}^\infty {\cal D} \bX_{r_1} ... {\cal D} \bX_{r_k} \psi^* (\bX_{r_1},...,\bX_{r_k})
  \left (\sum_{s=1}^k \bX_{r_s}  \right )\psi(\bX_{r_1},...,\bX_{r_k}) .
\lbl{2.72}
\ear
Explicitly, the first term in the above expression reads:
\be
   \int {\cal D}\, \bX_r \psi^* (\bX_r) \bX_r \psi (\bX_r)  = \sum_{r=1}^\infty \dd^3 \bx_1 ...\dd^3 \bx_r
   \psi^*(\bx_1,...,\bx_r) \sum_{s=1}^r \bx_s \, \psi (\bx_1,...,\bx_r) .
\lbl{2.73}
\ee
It corresponds to the expectation value (\ref{2.46}) of position in Quantization II.

In Quantization II, the expectation value of the position operator $\langle \hbx \rangle$ gives
the expected position of the effective single particle (the center of mass) associated with a superposition
of multiparticles states. Not only a single particle wave
packet profile $f(\bx)$, but also multiparticle wave packet profiles $f(\bx_1,...,\bx_r)$,
$r=1,2,...,\infty$ contribute to $\langle \hbx \rangle$.

In Quantization III, the expectation value of the configuration position operator
$\langle {\hat \bX}_r \rangle$ gives the effective $r$-particle configuration
associated with a superposition (\ref{2.66}) of multi configuration states.
They are determined not only with the $r$-particle wave packet profile
$\psi(\bX_r) \equiv \psi(\bx_1,\bx_2,..., \bx_r)$, but also with all other multi configuration
wave packet profiles, $\psi(\bX_{r_1},\bX_{r_2}), ...., \psi(\bX_{r_1},...,\bX_{r_s})$,
$s=1,2,...,\infty$.

As in Quantization II we do not have single particle states only, but in general
a multi particle state, so in Quantization III we have multi configuration states\footnote{
How successive quantization work for the energy eigenstates of a harmonic oscillator is
described by Baez\,\ci{Baez}. }
that comprise configurations of configurations according to Eq.\,(\ref{2.67}). This means
that  the predictions of the current solid state and particle physics based on quantum
field theories formulated within Quantization II will turn out not to be exactly valid,
if Quantization III is a valid physical theory. It predicts new phenomena---let
me call them ``miracle states'', or shortly,``miracles''---that are outside the reach of the current quantum mechanics
and quantum field theories. This opens
a pandora box of new possible directions of research.

\section{Discussion}

In quantum mechanics a wave function $\psi (t,\bx)$ evolves deterministically according to the Schr\"odinger
equation $i \p \psi(\bx)/\p t = h \psi(\bx)$, where $h=h(\bx,\p/\p \bx)$ is a Hamilton operator (\ref{2.14}),(\ref{2.15})
or (\ref{2.16}).
Upon observation, the wave function ``collapses'' into one of the eigenvalues
of the operator describing the measured quantity. Such a non deterministic evolution needs
a deeper explanation. We have seen that because the wave function itself can be
quantized, it in fact does not evolve deterministically, but can be at a given time $t$ any function $\psi (\bx)$,
including a function sharply localized around a position $\bx_0$, i.e., a
collapsed wave function.
The probability density of its occurrence is determined by a wave functional of $\Phi[\psi(\bx)]$, whose
absolute square $|\Phi[\psi(\bx)]|^2$ is the probability density of observing a function $\psi(\bx)$.
The wave functional $\Phi[\psi(\bx)]$ evolves according a determinist equation which is a higher level 
Schr\"odinger equation, $i\p \Phi[\psi(\bx)]/\p t = H[\psi(\bx),\frac{\delta}{\delta \psi (\bx)}] \Phi[\psi(\bx)]$,
where $H= \int \dd^3 \bx \,\psi(\bx)^* h(\bx,\frac{\p}{\p \bx}) \psi (\bx)$. Upon measurement of the
operator ${\hat \psi}(\bx)$, notwithstanding the problem of how such measurement can be done
in practice (however, see Ref.\,\ci{Lunden}),
we observe one of its eigenvalues, i.e., a definite function, say $\psi_0(\bx)$.
This means that the functional $\Phi[\psi(\bx)]$ collapses into a functional which is sharply localized
around a definite function $\psi_0 (\bx)$. And this function $\psi_0(\bx)$ can be {\it any function}, the
probability (density) of its occurrence being given by $|\Phi[\psi(\bx)]|^2$. Upon measurement of some
other observable, say position ${\hat \bx}$, or momentum $\hbp$, we observe that their expectation values
deviates from the usual ones,  $\langle \hbx \rangle = \int \dd^3 \bx \,\psi^*(\bx) \bx \psi (\bx)$,
or $\langle \hbp \rangle = \int \dd^3 \bp \,\psi^*(\bp) \bp \psi (\bp)$,  as shown in Eq.\,(\ref{2.72}),(\ref{2.73}),
if we take into account that $\psi(\bx)$ is quantized. A next step, namely  the quantization of
the wave functional $\Phi[\psi (\bx)]$, was also discussed in this paper.

In our development of the formalism of subsequent quantizations, we split the wave function into its
real and imaginary part, $\psi = \psi_R + i \psi_I$,
and considered the second order equation equation ${\ddot \vphi}(\bx) + h^2 \vphi(\bx) =0$
for $\vphi(\bx) \equiv \psi_R(\bx)$. We then quantized $\vphi(\bx)$ and obtained the first
order Schr\"odinger equation  for the complex wave functional $\Phi[\vphi(\bx)]=\Phi_R + i \Phi_I$, or, equivalently,
the second order equation ${\ddot \chi}[\vphi] + H^2 \chi[\vphi] =0$
for its real part $\chi[\vphi] \equiv \Phi_R [\vphi]$. We then also quantized $\chi[\vphi]$ and arrived at the
next level Schr\"odinger equation. This last step was in fact the quantization of the usual quantum field theory
in which states are given in terms of the multiparticle wave packet profiles. And the latter profiles have been quantized which leads to deviations from the prediction of quantum field theory.

Finally let us mention that although we stopped at the level of Quantizatoin III,
the process of successive quantizations can continue ad infinitum according to $\bx$, $\vphi^{(1)}(\bx)$, $\vphi^{(2)}[\vphi^{(1)}(\bx)]$,
$\vphi^{(3)}[\vphi^{(2)}[\vphi^{(1)}(\bx)]]$,..., $\vphi^{(n)}[\vphi^{(n-1)}[\vphi^{(n-2)},...., \vphi^{(1)}(\bx)]]$,...\,.
Given the fact that mathematics was
very successful in treating the infinite iterative procedures, such as the integration, infinite series, etc.,
we may anticipate that also the process of the $n$-th quantization in the limit of infinite $n$ can be treated in
a similarly successful manner. The first step, namely the quantization of wave function can be understood 
as an explanation of the wave function collapse. But on second thought we find the the enigma of collapse
is only postponed or shifted to the next level of quantization, where the wave functional upon measurement
collapses so to be picked on one of many possible wave functions. And so on along the chain of $n$-th quantizations.
It would interesting to find out what in the limit of infinite $n$ happens with wave function(al) and its collapse on the one hand,
and with its many worlds interpretation\,\ci{Everett} on the other hand.


\begin{thebibliography}{12}

\bi{Dirac}P. A. M. Dirac, {\it Proc. Roy. Soc. London A}  {\bf 114}, 243 (1927).

\bi{Jordan} P. Jordan, E. Wigner, {\it Z. Phys.} {\bf 47}, 631 (1928).

\bi{Fock}  V. A.  Fock,   Z. Phys. {\bf 75}, 622 (1932).

\bi{Hatfield} B. Hatfield, {\it Quantum Field Theory of Point Particles and Strings} (Adison-Wesley,  Reddwood City,1992).

\bibitem{PavsicManifestCovariance} 
  M. Pav\v si\v c,
  {\it Mod.\ Phys.\ Lett.\ A} {\bf 33}, no. 20, 1850114 (2018)\\
  doi: 10.1142/S0217732318501146
  [arXiv:1804.03404 [hep-th]].
  %%CITATION = doi:10.1142/S0217732318501146;%%
 
\bi{PavsicLocalized} M. Pav\v si\v c,
%``Localized States in Quantum Field Theory,''
  {\it Adv.\ Appl.\ Clifford Algebras} {\bf 28}, no. 5, 89 (2018)\\
  doi: 10.1007/s00006-018-0904-5
  [arXiv:1705.02774 [hep-th]].
  %%CITATION = doi:10.1007/s00006-018-0904-5;%%

\bi{Lunden}  J.S. Lunden, B. Sutherland, A. Patel, C. Stewart and C. Bamber, {\it Nature} {\bf 474}, 188 (2011).

\bi{Deriglazov} A. A. Deriglazov, {Phys. Lett. A} {\bf 373}, 3920 (2009), 10.1016/j.physleta.2009.08.050 [arXiv:0903.1428 [math-ph]].
\bi{Foldy} L. L. Foldy, {\it Phys. Rev.} {\bf 102}, 568 (1956). 

\bi{Baez} J. Baez, {\it The Story of Nth Quantization}\\
http://math.ucr.edu/home/baez/nth\_quantization.html

\bi{Everett} H. Everett, {\it Rev. Mod. Phys} {\bf 29}, 454 (1957).

\end{thebibliography}
\end{document}